\documentclass{ws-mpla} \usepackage{graphicx}

\newcommand{\xmax}{\ensuremath{X_{\rm max}}}
\newcommand{\nmax}{\ensuremath{N_{\rm max}}}

\begin{document}

\markboth{Douglas~R.~Bergman} {Chasing the GZK with HiRes}

%
\catchline{}{}{}{}{}
%

\title{CHASING THE GZK WITH HiRes\footnote{Expanded version of a
    seminar given at Pennsylvania State University on 23 October
    2002.}}

\author{\footnotesize DOUGLAS R. BERGMAN}

\address{Rutgers - The State University of New Jersey, Department of
  Physics and Astronomy, Piscataway, NJ, USA\\
  bergman@physics.rutgers.edu}

\maketitle

\pub{Received (Day Month Year)}{Revised (Day Month Year)}

\begin{abstract}
  The HiRes Collaboration has recently announced preliminary
  measurements of the energy spectrum of ultra-high energy cosmic rays
  (UHECR), as seen in monocular analyses from each of the two HiRes
  sites.  This spectrum is consistent with the existence of the GZK
  cutoff, as well other aspects of the energy loss processes that
  cause the GZK cutoff.  Based on the analytic energy loss formalism
  of Berezinsky {\it et al.}, the HiRes spectra favor a distribution
  of extragalactic sources that has a similar distribution to that of
  luminous matter in the universe, both in its local over-density and
  in its cosmological evolution.  
\end{abstract}

\ccode{PACS Nos.: 95.85.Ry, 96.40.De, 96.40.Pq, 98.70.Sa}

\section{Introduction}

The cosmic ray energy spectrum is nearly featureless over ten orders
of magnitude in energy, from $10^{10}$ eV to $10^{20}$ eV, with the
differential flux falling approximately as $E^{-3}$.  There are three
small, though widely discussed, features: the ``knee'', a hardening of
the spectrum at $10^{15.5}$ eV; the ``second knee'', another hardening
at about $10^{17.6}$ eV; and the ``ankle'', a softening of the
spectrum at about $10^{18.5}$ eV.  These features may represent
changes in the sources, composition or dynamics of the cosmic rays.
Two often asked questions are: How do cosmic rays come to have such
high energies (a joule or more of kinetic energy in a proton or other
sub-atomic particle), and does the spectrum continue above $10^{20}$
eV?

There are two types of models describing the sources of ultra-high
energy cosmic rays (UHECRs): astrophysical models (``bottom-up''), in
which cosmic rays are accelerated to very high energies by magnetic
shock fronts moving though plasmas; and cosmological models
(``top-down''), in which the cosmic rays are the result of the decays
of super heavy particles which are relics of the Big Bang.  I will
only be discussing the former.  One can evaluate the plausibility of
various astrophysical sources by considering the magnetic field of the
object and its size.\cite{Hillas-84} The overall magnetic field
contains the nascent cosmic rays during their acceleration and thus
must be large enough to keep the cosmic rays within the object.
Smaller objects need larger fields; larger objects, smaller fields.
By this criterion we have several candidate sources: neutron stars,
active galactic nuclei (AGN) and clusters of galaxies among others.
All these sources could plausibly, by the above argument, give cosmic
rays at $10^{20}$ eV, but, in all cases, one is pushing the bounds of
plausibility at the highest energies.

If UHECRs are extragalactic, then they must traverse the intergalactic
medium in order to be observed.  This medium is filled with cosmic
microwave background (CMB) photons, which should lead to a fourth, and
not so small, feature of the UHECR spectrum.  Because of their large
kinetic energies, UHECRs interact with the CMB to produce resonances
(in the case of protons) or to dissociate (in the case of nuclei).  In
the proton case, the resonance (e.g. $\Delta^+$) will decay quickly
into proton or neutron and a meson (e.g. $\pi$).  In either case, the
result is a reduction in the energy of the leading particle.  At
somewhat lower energies, cosmic rays lose energy by creating
electron-positron pairs in their interaction with the CMB.  These
energy loss mechanisms imply that there should be a sharp reduction in
the UHECR flux above $10^{19.8}$ eV, assuming the UHECRs are
protons and that they come from distances greater than a few tens of
megaparsecs.  Nuclei should have an even lower energy threshold.  This
fact, first pointed out by Greisen, Zatsepin and Kuzmin, has become
known as the GZK cutoff.\cite{Greissen-66,Zatsepin-66} By measuring
the shape of the UHECR spectrum and, crucially, modeling the spectrum
at the source, one can hope to deduce which of the plausible sources
listed above, if any, contribute to the UHECRs we see.

If UHECRs are produced in our galaxy they are not subject to the GZK
cutoff.  However, there are no plausible astrophysical accelerators of
UHECRs within our galaxy.  Any such object would appear as a point
source in a map of the sky made with UHECRs, due to the short
propagation distances and relatively weak magnetic fields.  No such
point source has been observed.

\section{Experimental Techniques}

UHECRs have a very low flux, so one must have a large collection area
to obtain a reasonable event rate.  This precludes direct observations
of UHECR above the Earth's atmosphere in satellite experiments.
However, one may also use that atmosphere as a giant calorimeter,
because UHECRs create extensive air showers (EASs) when they encounter
the atmosphere.  This allows access to very large areas.

There are two ways to instrument this atmospheric calorimeter: readout
the particle multiplicities at the back end by putting arrays of
detectors on the ground, or collect the light produced as the EAS
gives up its energy to the atmosphere.  The former technique (Ground
Arrays) has the advantage of 100\% duty cycle: one can run at all
times of the day.  It has the disadvantage that one usually observes
only the tail end of the EAS and has to infer the properties of the
primary particle rather indirectly.  To illustrate, consider the
lead-scintillator sandwich type calorimeter used in many fixed-target
experiments at accelerators.  One normally collects the light produced
by the shower as it goes through the scintillator segments.  The total
light is proportional to the energy of the initial particle, and one
can in principle measure the longitudinal development of the shower.
Now imagine throwing away the signals from all but the last
scintillator segment and one can understand the difficulties faced by
ground arrays.  One must also make a trade-off between density of
detectors on the ground and the total area over which one places
detectors.

Collecting the fluorescence light from EASs has complementary
advantages and disadvantages.  The main advantage is that one observes
light from all stages in the development of the EAS, and the amount of
this light is directly proportional to the primary energy.  The
disadvantage is that one is subject to the optical changes inherent in
the atmosphere and one can only run when and where it is dark and
clear.  As a counterpart to the example above, fluorescence detectors
are like lead-glass calorimeters, where one collects light from the
whole detector element.  However, the glass may be somewhat smoky.

The Akeno Giant Air Shower Array (AGASA)\cite{Sakaki-01} is the
largest, currently active example of a ground array.  The AGASA
collaboration claims to see no evidence for the GZK
cutoff,\cite{AGASA-spec} which has motivated a great deal of
theoretical work on possible mechanisms by which the GZK cutoff could
be avoided.  The Fly's Eye Experiment\cite{Baltrusaitis-85} is an
example of a fluorescence detector, and the experiment that has
observed the highest energy cosmic ray ever detected at
$3\times10^{20}$ eV.\cite{Bird-93} The Pierre Auger
Observatory,\cite{Auger} currently under construction, will combine
both a very large ground array and a fluorescence detector, in an
effort to have the advantages of both types of detectors.

\section{The HiRes Detector}

The High Resolution Fly's Eye Experiment (HiRes) is a direct
descendant of Fly's Eye, designed with bigger mirrors and finer
pixels, to give a larger aperture by a factor of ten.  It consists of
two sites, separated by 12 km, in order to observe EASs in stereo.
Stereo observation greatly reduces the uncertainty in the geometrical
reconstruction of the EAS.  The sites are located on hills on the
Dugway Proving Grounds in the west desert of Utah.  The remote desert
provides a dark, optically clean atmosphere, while the hills put the
detectors above much of the remaining aerosols.

Each detector consists of mirror units viewing a
$14^\circ\times16^\circ$ patch of the sky with 256 photomultiplier
tubes (PMTs), each of which views about $1^\circ$, in a $16\times16$
array.  Each mirror has an area of about 5 m$^2$.  The HiRes-I site,
the first of the two to be built, has one ring of mirrors covering
from $3^\circ$--$16^\circ$ and nearly the complete azimuth.  The PMTs
are read out using a sample-and-hold technique, that gives the time
and size of the signal for each tube.  The HiRes-II site has two rings
of mirrors covering $3^\circ$--$30^\circ$.  These PMTs are read out
using a flash ADC (FADC) system, which samples each of the tubes every
100 ns.  This provides the shape of the signal in each tube and allows
one to combine the light from different tubes that were active at the
same time.  HiRes-I began operation in June of 1997.  HiRes-II began
in October of 1999.

\section{Monocular Analyses}

The reader is referred to the published Fly's
Eye\cite{Baltrusaitis-85} and HiRes\cite{AbuZayyad-02a,AbuZayyad-02b}
papers for details of the reconstruction techniques.  Only a brief
summary will be given here.

\subsection{EAS Geometry}

Although HiRes was designed as a stereo experiment, there are two
reasons for continuing to consider monocular analyses.  First, since
HiRes-I was running for two years before HiRes-II came on-line, the
largest UHECR data sample is the HiRes-I monocular sample.  Second,
low-energy events are close to one or the other of the two sites, and
trigger that site only.  Thus, the low-energy reach of the detector
will always be in monocular mode.

HiRes-II is a better detector for reconstructing monocular events, due
to its two rings: longer tracks lead to a better determination of the
EAS geometry.  There are two tasks in determining the geometry of an
EAS: finding the shower-detector plane (SDP) and determining the angle
of the shower within the SDP.  The geometry of the shower within the
SDP is determined by fitting the time of the tube signals
\begin{equation}
t_i = t_0 + \frac{R_p}{c} \tan\left(\frac{\pi-\psi-\chi_i}{2}\right)
\label{eq:tvsa}
\end{equation}
for $R_p$, $\psi$ and $t_0$, where $t_i$ is the signal time in the
$i$th tube, $R_p$ is the impact parameter, $t_0$ is the time the
shower core reaches the $R_p$ point, $\psi$ is the angle of the EAS in
the SDP and $\chi_i$ is the viewing angle in the SDP of the $i$th
tube.  Longer tracks make it easier to distinguish the tangent
function from a straight line.  HiRes-I tracks are often too short to
resolve all the ambiguities from timing alone, and one must look to
the reconstructed shower profile (see below) for assistance in
determining the geometry.

\begin{figure}
  \begin{center}
    \includegraphics[width=0.49\columnwidth]{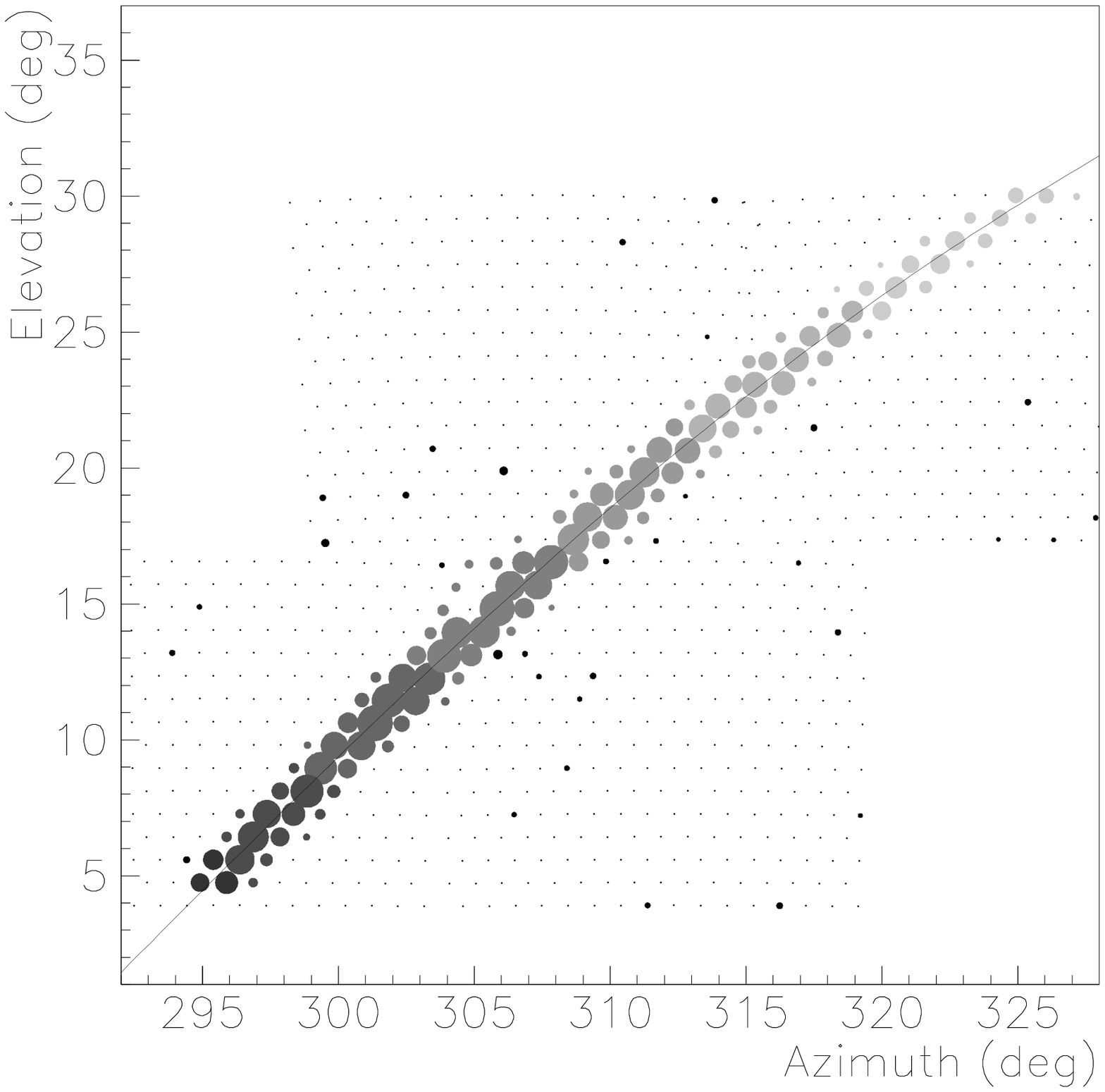}
    \includegraphics[width=0.49\columnwidth]{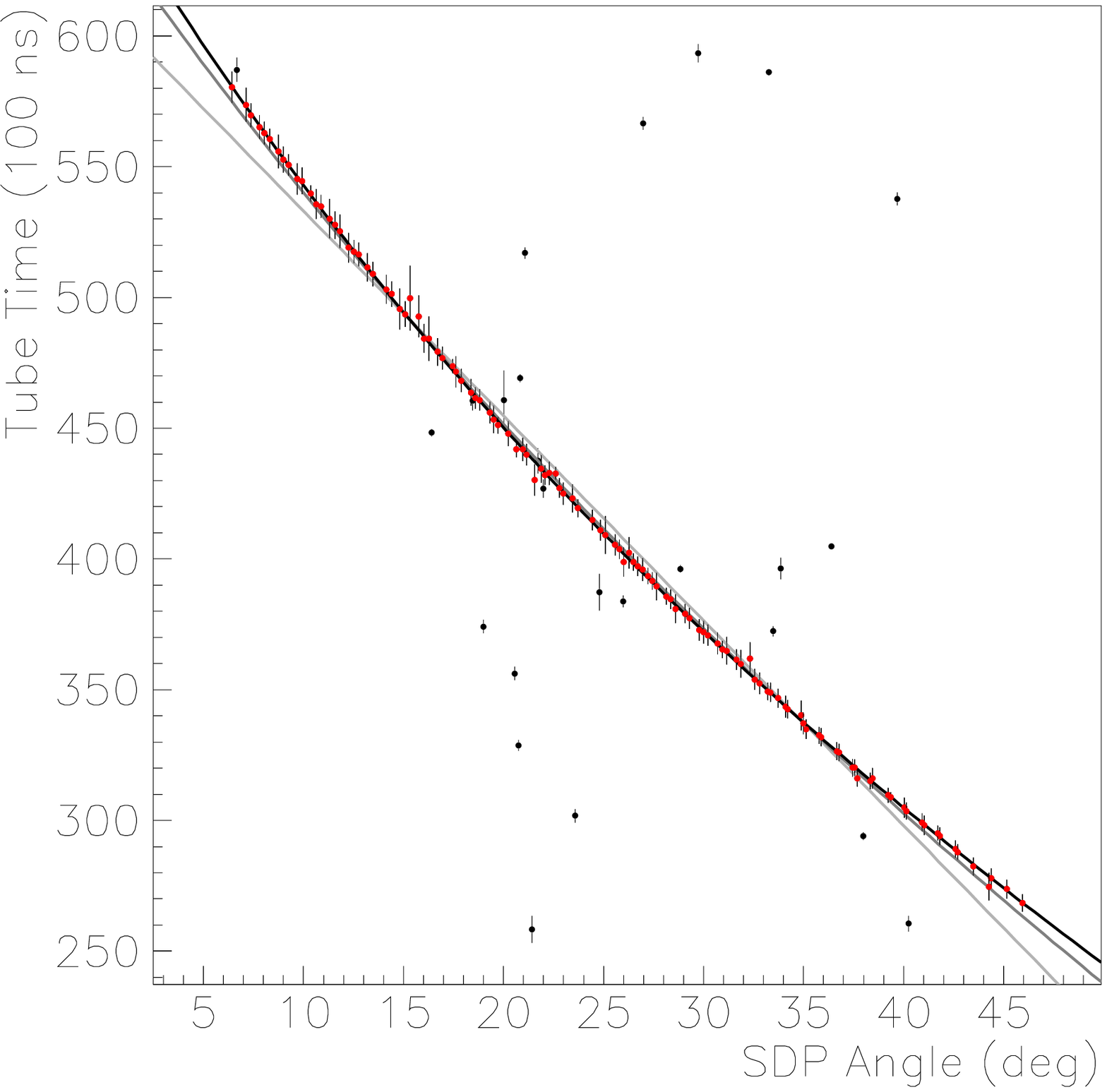}
  \end{center}
  \vspace*{8pt}
  \caption{A 50 EeV (1 EeV = $10^{18}$ eV) UHECR event, as seen on the
    mirrors/PMTs of the detector (left) and in a time vs (SDP) angle
    plot (right).  In the mirror/PMT display dots represent inactive
    tubes, circles represent active tubes where the size of the tube
    is proportional to the signal.  Shaded tubes are included in the
    fit, with the shading representing the relative times: light to
    dark, early to late.  In the time vs angle plot, the time is shown
    in units of 100 ns.  The three fits are from Eq~\ref{eq:tvsa} with
    $\psi=180^\circ$ (light grey), $\psi=90^\circ$ (dark grey) the
    best fit $\psi$ (black).}
  \label{fig:mirror-tvsa}
\end{figure}

As an example, a picture of a 50 EeV cosmic ray event from HiRes-II,
given in Fig.~\ref{fig:mirror-tvsa}, shows the azimuthal and elevation
angles of all the tubes in the mirrors that were part of the event.
Inactive tubes are shown as dots; active tubes are shown as filled
circles, where the radius is proportional to the tube signal.  Active
tubes that are used in fitting the SDP are shaded according to the
average time of the FADC measurements of the tube.  The fitted SDP is
also shown in the figure.  The average time of the signal for each
tube as a function of the angle ($\chi_i$) in the SDP is also shown
for the same event, including three fits to Eq.~\ref{eq:tvsa}, one
with $\psi=180^\circ$ (light grey), one with $\psi=90^\circ$ (dark
grey) and the best fit $\psi$ (black).

\subsection{Calibration}

Once the geometry of the EAS is determined, one can use the
photoelectron (NPE) signal of the PMTs to determine the number of
charged particles, $N_e$, in the shower.  In making this ${\rm NPE}
\rightarrow N_e$ conversion, two important calibration issues come
into play: the gains of the PMTs, and the atmospheric transparency to
light.  PMT gains are monitored nightly and monthly using a Xenon
flash lamp and YAG laser.  The atmosphere is monitored using a
bistatic LIDAR system.

\subsection{Fitting Profiles and Determining Energy}

The HiRes-II analysis combines the PE signal from all the tubes in a
given time bin, and converts this into a given number of charged
particles in the EAS at a given depth in the atmosphere (measured in
g/cm$^2$).  This conversion is strongly dependent on the geometry.
One can estimate the energy of a shower from the number of charged
particles, the amount of material traversed, and the average energy
deposited per particle:
\begin{equation}
  E_0 = 2.19 {\rm\ MeV/(g/cm^2)} \int^\infty_0 N_e(X) dX
\end{equation}
Since one often does not see the entire EAS, one must assume some
profile of the number of charged particles for the unobserved part of
the shower.  We use the Gaisser-Hillas parameterization
\begin{equation}
  N_e(X) = \nmax
  \left(\frac{X-X_0}{\xmax-X_0}\right)^\frac{\xmax-X_0}{\lambda}
  \exp\left(\frac{X-\xmax}{\lambda}\right)
\end{equation}
where $\xmax$ is the depth at the maximum extent of the EAS, $\nmax$
is the number of particles at that depth, $X_0$ corresponds to the
depth of the first interaction, and $\lambda$ is the interaction
length.  We fitted the observed portion of the shower to determine
$\xmax$ and $\nmax$, holding $X_0=-60$ g/cm$^2$ (which is not
physical, but gives the best fits when applied to simulations using
CORSIKA\cite{CORSIKA}) and $\lambda=70$ g/cm$^2$.

\begin{figure}
  \begin{center}
    \includegraphics[width=0.75\columnwidth]{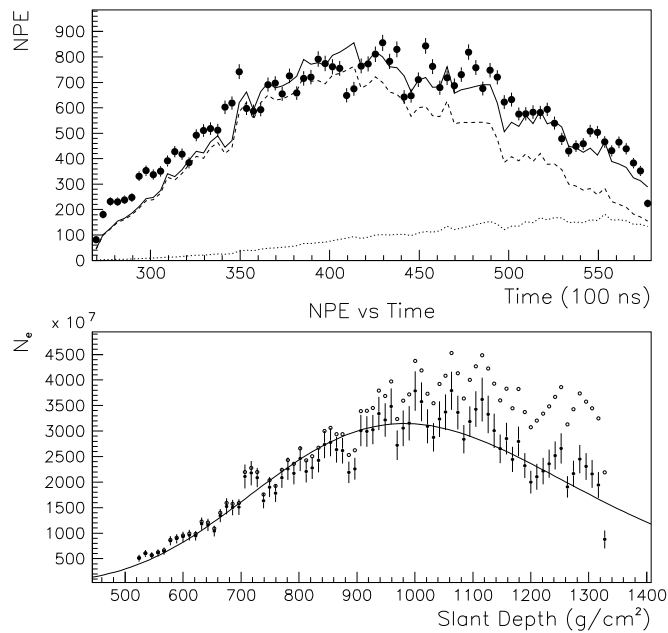}
  \end{center}
  \vspace*{8pt}
  \caption{NPE and $N_e$ profiles in the 50 EeV event of
    Fig.~{fig:mirror-tvsa}.  In the the upper plot, points with error
    bars show the photoelectron signal (NPE) from all tubes included
    within a given time bin.  The dashed and doted lines show
    calculations for the fluorescence and \v{C}erenkov components of
    the signal, respectively, for this event at the best fit values of
    $\xmax$ and $\nmax$; the solid line shows the sum of the
    components.  In the lower plot, filled points with error bars show
    the calculated $N_e$ extracted from the NPE signal as a function
    of depth in the atmosphere.  The open points without error bars
    show the result of the calculation without subtracting off the
    \v{C}erenkov component.}
  \label{fig:profile}
\end{figure}

Some of the observed light comes from the beam of \v{C}erenkov light
generated by the EAS and scattered into the detector.  This light is
subtracted from the signal in an iterative procedure.  The
photoelectron signal as a function of time and the calculated $N_e$ as
a function of depth for the same 50 EeV event are shown in
Fig.~\ref{fig:profile}.

The energy is calculated from the HiRes-I signals in a similar way,
except that the expected number of PE for a given shower is compared
with the observed value.  In other words, the fit is done using the
number of PE at the detector rather than the extracted number of
charged particles in the shower.  The comparison is also done on a
tube-by-tube basis rather than in time bins.

\subsection{Event Selection}

After events are reconstructed, one must select the sample from which
to calculate the flux.  This sample should be as large as possible and
contain only well reconstructed events.  The criteria used to make
this selection are listed in Table~\ref{tab:cuts}.

\begin{table}[h]
\tbl{Cuts used in HiRes-I and HiRes-II monocular analysis flux
  calculations.}
{
  \begin{tabular}{@{}lrr@{}}\toprule\toprule
                                   & HiRes-I   & HiRes-II     \\ \toprule
      Minimum distance (Ang. Vel.) & 5  km     & 1.5 km       \\
      Minimum track length         & 8$^\circ$ & 10$^\circ$   \\
      \ \ \ \ \ in ring 1 only     &           &  7$^\circ$   \\
      Range of tubes per degree    & [0.85,3]  & [0.85,3]     \\
      Minimum NPE per degree       & 25        & 25           \\
      Maximum zenith angle         &           & 60$^\circ$   \\
      Minimum depth seen           &           & 150 g/cm$^2$ \\
      Minimum extent seen          &           & 150 g/cm$^2$ \\
      $X_{\rm max}$ observed       &           & required     \\
      Maximum \v{C}erenkov correction &        & 60\%         \\
      Profile contraint converges  & required  &              \\ 
      Data period starting         & 6/1997    & 12/1999      \\
      Data period ending           & 9/2001    & 5/2000 \\
      \toprule\toprule
  \end{tabular}
}
\label{tab:cuts}
\end{table}

\section{Monte Carlo Simulation}

To determine the flux of UHECR, one needs to know the aperture over
which the UHECRs are collected.  This aperture varies with energy and
must be determined through Monte Carlo (MC) simulation.  However, one
can also use MC simulation to check that one understands the data and
its reconstruction in the detector.  Extensive comparisons of
distributions in data and in simulated samples of events provides
confidence that the calculated aperture is correct.

The details of MC event generation have been published
elsewhere.\cite{AbuZayyad-02b} It is clear, however, that one needs to
model the details of the trigger, the extra tube distribution, and the
transmission of light in the atmosphere to sufficient accuracy to
obtain good agreement between data and MC.

\begin{figure}
  \begin{center}
    \includegraphics[width=0.49\columnwidth]{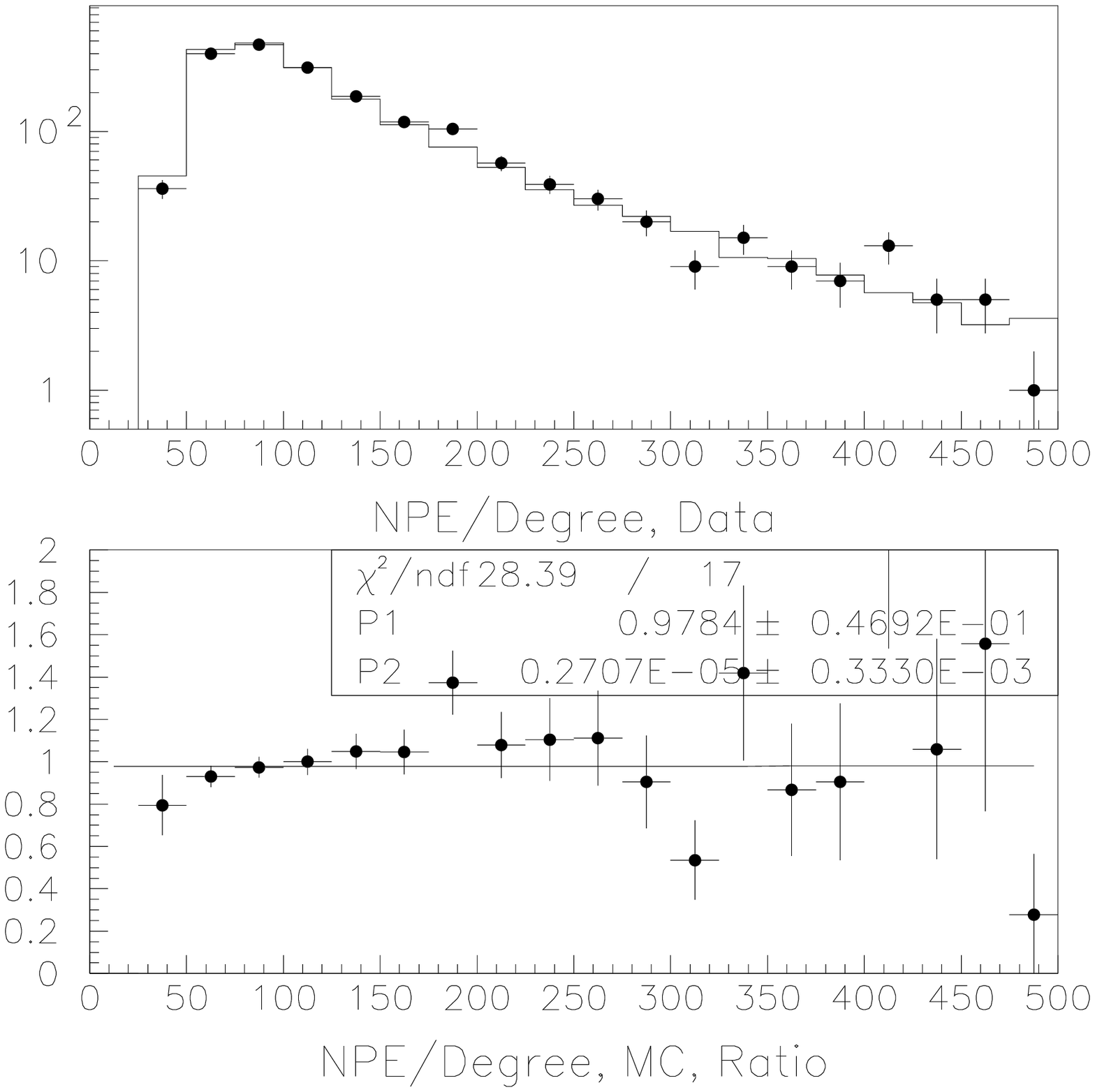}
    \includegraphics[width=0.49\columnwidth]{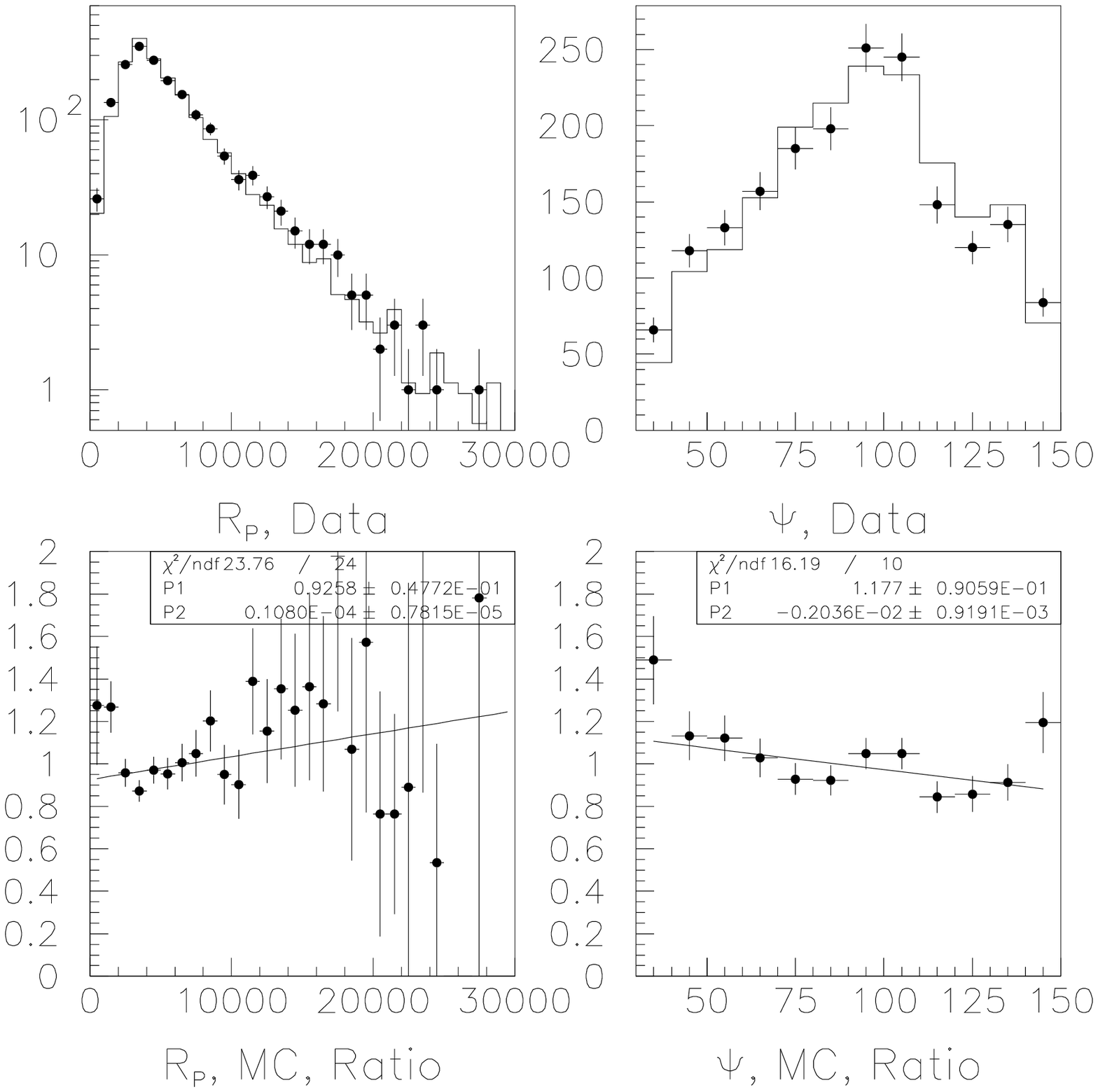}
  \end{center}
  \vspace*{8pt}
  \caption{Comparisons of the number of photoelectrons (NPE) per
    degree of track (left) and the EAS geometry variables $R_p$ and
    $\psi$ (right) between data (points) and MC (histogram) where in
    each case the MC distribution has been normalized to have the same
    area as the data.  The bottom plots show the ratio of the two
    above distributions: data/MC.}
  \label{fig:mcnpe-geo}
\end{figure}

\begin{figure}
  \begin{center}
    \includegraphics[width=0.75\columnwidth]{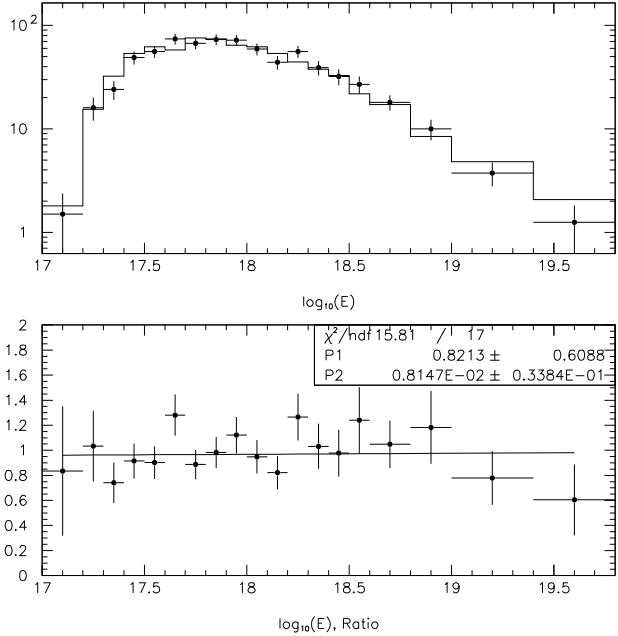}
  \end{center}
  \vspace*{8pt}
  \caption{A comparison of the reconstructed energy between data
    (points) and MC (histogram), where the MC distribution has been
    normalized to have the same area as the data.  The bottom plot
    shows the ratio of the two distributions: data/MC.  The vertical
    scale is number of events per 0.1 decade of energy.}
  \label{fig:mcenergy}
\end{figure}

\begin{figure}
  \begin{center}
    \includegraphics[width=0.75\columnwidth]{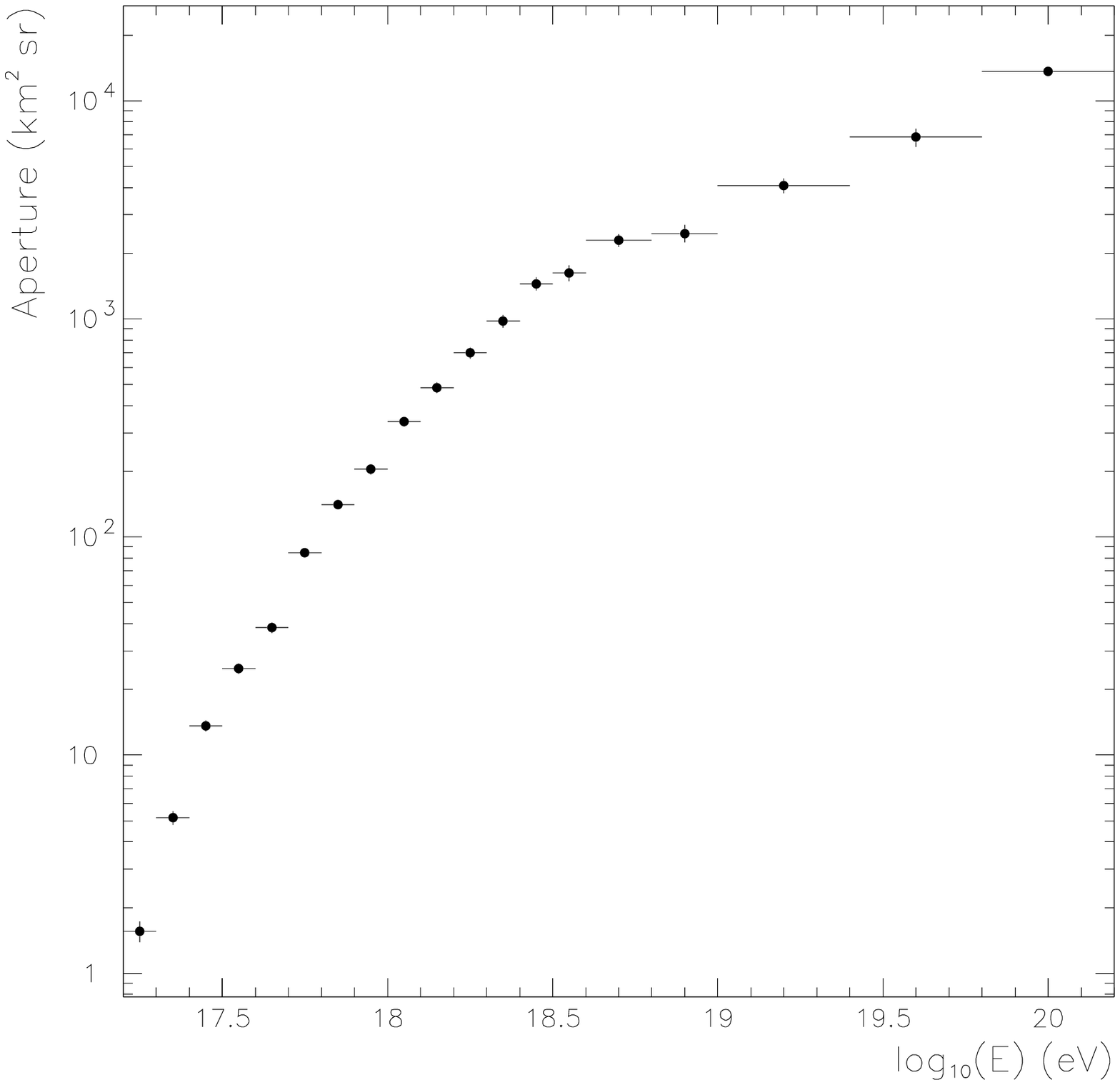}
  \end{center}
  \vspace*{8pt}
  \caption{The HiRes-II aperture as calculated in the MC.}
  \label{fig:aperture}
\end{figure}

I will show four comparisons between data and MC, all taken from the
HiRes-II analysis.  The first, in Fig.~\ref{fig:mcnpe-geo}, shows the
distribution of observed light (NPE) divided by the track length.  The
light distribution is sensitive to the yield, trigger, and geometry,
among other things.  The second and third comparisons, also in
Fig.~\ref{fig:mcnpe-geo}, show the distributions of $R_p$ and $\psi$.
The good agreement gives us confidence in our calculation of the
aperture.  Finally, there is the energy distribution in
Fig~\ref{fig:mcenergy}, which enters directly into the calculation of
the flux.  Note that the energy distribution has a binning such that
there are no bins with less than two events.  The HiRes-II aperture,
with the same binning, is shown in Fig.~\ref{fig:aperture}.

\section{Flux}

\begin{figure}
  \begin{center}
    \includegraphics[width=0.75\columnwidth]{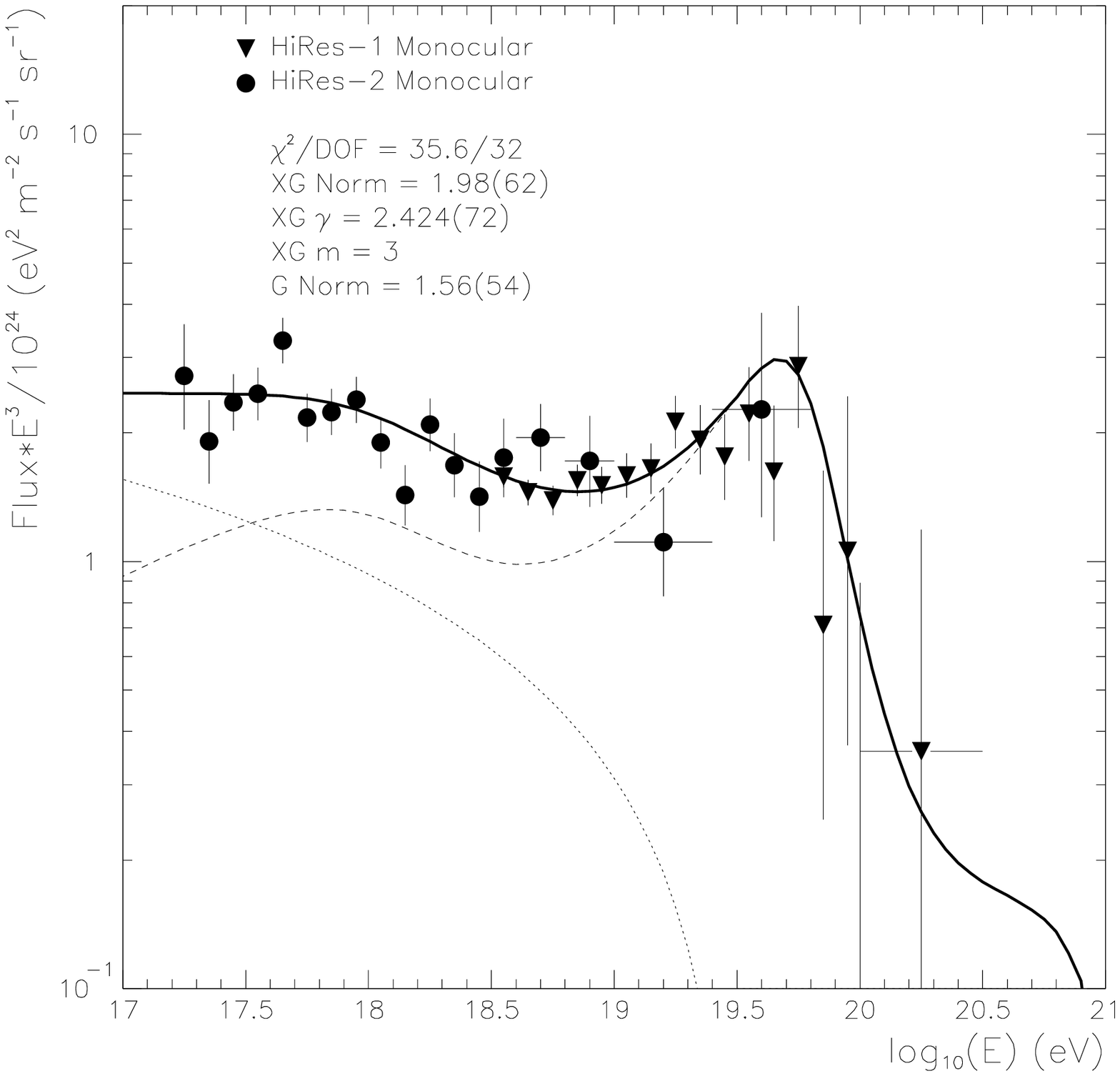}
  \end{center}
  \vspace*{8pt}
  \caption{The UHECR flux, multiplied by $E^3$, as measured by HiRes-I
    (triangles) and HiRes-II (circles).  The solid line is a two
    component fit to the data consisting of a galactic (dotted line)
    and extragalactic (dashed line) spectrum.  The extragalactic
    sources evolve as $(1+z)^m$ with $m=3$ and have a distribution
    modified by the observed density of galaxies.}
  \label{fig:fluxm3}
\end{figure}

With the event samples in hand, and confidence in our calculated
aperture, we can extract the flux.  Fig.~\ref{fig:fluxm3}, shows the
calculated flux from HiRes-I (triangles) and HiRes-II (circles).  The
fluxes have been multiplied by $E^3$ in order to emphasize changes in
the spectral index.

To evaluate the significance of these spectra, I have fitted the
measured points to an astrophysical model of the sources and
propagation of UHECRs.  In choosing this model, I have been motivated
by Occam's Razor, sticking with known physical processes, and assuming
only that the extragalactic sources of UHECR are distributed
throughout the universe, and evolve in their density in the same way
as the luminous matter in galaxies.  To this is added a
phenomenologically motivated galactic spectrum at lower energies.

The extragalactic spectrum is assumed to consist of protons, and have
a power law spectrum at the source, with a fitted spectral slope
parameter.  This spectrum is modified by energy losses as the UHECRs
traverse the intra-galactic medium.  The energy loss formalism is
taken from the work of Berezinsky {\it et
al}.\cite{Berezinsky-88,Berezinsky-02} The sources are taken to be
uniformly distributed out to a red-shift of $z=4$, with a density at
any given $z$ modified by the observed density of galaxies at that
red-shift\cite{Blanton-01,Saunders-00} and evolving as $(1+z)^m$ with
$m=3$ (which is the best fit value and approximately the same as the
observed stellar formation rate\cite{Lilly-96}).  Using $m=0$,
increases the $\chi^2$ of the fit by 3.5 while not accounting for the
observed density of galaxies increases the $\chi^2$ by 1.5.

The galactic component of the spectrum is assumed to consist of iron
nuclei.  Motivation for this assumption comes from the Fly's Eye
composition measurement,\cite{Bird-95} which shows an approximately
linear (in $\log E$) change from a heavy composition (iron) to a light
one (protons).  The spectral form is assumed to be $E^{-3}$,
consistent with the UHECR spectrum below $10^{17}$ eV, multiplied by a
linear (in $\log E$) factor going from unity at $10^{17}$ eV, to zero
at $10^{19.5}$ eV.

The fact that this model fits the data so well is due to agreement in
its three features.  First, the model clearly has a drastic reduction
in flux above the GZK energy, a reduction which is also observed in
the data, but with not as yet overwhelming statistical significance.
More strongly constrained by observation are the ``second knee'' and
``ankle'', which in this model are the result of electron pair
production.\cite{Berezinsky-88,Berezinsky-02} The strength of the
astrophysical source model is that it fits both the electron
pair-production features and the pion production feature (the GZK
cutoff) in the spectrum.

\section{Conclusion}

HiRes has recently announced preliminary measurements of the UHECR
energy spectrum using each of its two detectors in monocular mode.  We
have fitted this spectrum to a two-component model of the UHECR
sources, the extragalactic component of which conforms to the
expectation of a nearly unform distribution of sources, modified by
the observed distribution of luminous matter and its evolution as a
function of red-shift.  This model fits all the recognized features of
the UHECR spectrum, not just the GZK cutoff.

Much has been made of the discrepancies between HiRes and AGASA, with
AGASA seeing no evidence for the GZK cutoff, and HiRes being
consistent with its presence.  However, at energies below the GZK
cutoff, the two experiments agree nicely in the shape of the spectrum,
with only an offset in the flux, which can be attributed to a
difference in energy scale within the stated systematic uncertainty of
either experiment.\cite{Bergman-02}

\section*{Acknowledgments}

While comments about the conformity of the HiRes spectra with the GZK
cutoff are my own opinions, the measurement of the spectra themselves
would not have been possible without my colleagues in the HiRes
Collaboration, to whom I feel indebted for the realization of this
work.  This work is supported by the US NSF grant PHY 0073057.

\end{document}